\documentclass[a4paper,fleqn]{cas-sc}

\usepackage[numbers]{natbib}
\usepackage{amsmath,amsfonts,amssymb}
\usepackage{algorithm}
\usepackage{mathtools}
\usepackage{array}
\usepackage{algpseudocode}
\usepackage{textcomp}
\usepackage{stfloats}
\usepackage{url}
\usepackage{verbatim}
\usepackage{graphicx}
\usepackage{subcaption}
\def\tsc#1{\csdef{#1}{\textsc{\lowercase{#1}}\xspace}}
\tsc{WGM}
\tsc{QE}
\tsc{EP}
\tsc{PMS}
\tsc{BEC}
\tsc{DE}

\begin{document}
\let\WriteBookmarks\relax
\def\floatpagepagefraction{1}
\def\textpagefraction{.001}

\shorttitle{A Sysmon Incremental Learning System for Ransomware Analysis and Detection}

\shortauthors{Ispahany et~al.}

\title [mode = title]{A Sysmon Incremental Learning System for Ransomware Analysis and Detection}                      



%
\author[1,3]{Jamil Ispahany}[type=editor,
                        auid=000,bioid=1,
                        orcid=0000-0001-8224-2924]



\ead{Jispahany@csu.edu.au}



\affiliation[1]{organization={Charles Sturt University}, 
    city={Bathurst},
    state={NSW},
    postcode={2795},
    country={Australia}}

\author[2,3]{MD Rafiqul Islam}

\author[1,3]{M. Arif Khan}


\affiliation[2]{organization={Charles Sturt University}, 
    city={Albury-Wodonga},
    state={NSW},
    postcode={2640},
    country={Australia}} 

\author%
[1,3]
{MD Zahidul Islam}

\affiliation[3]{organization={Cyber Security Cooperative Research Centre}, 
    city={Kingston},
    state={ACT},
    postcode={2600},
    country={Australia}}

\cortext[cor1]{Corresponding author}



\begin{abstract}
In the face of increasing cyber threats, particularly ransomware attacks, there is a pressing need for advanced detection and analysis systems that adapt to evolving malware behaviours. Throughout the literature, using machine learning (ML) to obviate ransomware attacks has increased in popularity. Unfortunately, most of these proposals leverage non-incremental learning approaches that require the underlying models to be updated from scratch to detect new ransomware, wasting time and resources. This approach is problematic because it leaves sensitive data vulnerable to attack during retraining, as newly emerging ransomware strains may go undetected until the model is updated. Furthermore, most of these approaches are not designed to detect ransomware in real-time data streams, limiting their effectiveness in complex network environments. To address this challenge, we present the Sysmon Incremental Learning System for Ransomware Analysis and Detection (SILRAD), which enables continuous updates to the underlying model and effectively closes the training gap. By leveraging the capabilities of Sysmon for detailed monitoring of system activities, our approach integrates online incremental learning techniques to enhance the adaptability and efficiency of ransomware detection. The most valuable features for detection were selected using the Pearson Correlation Coefficient (PCC), and concept drift detection was implemented through the ADWIN algorithm, ensuring that the model remains responsive to changes in ransomware behaviour. We compared our results to other popular techniques, such as Hoeffding Trees (HT) and Leveraging Bagging Classifier (LB), observing a detection accuracy of 98.89\% and a Matthews Correlation Coefficient (MCC) rate of 94.11\%, demonstrating the effectiveness of our technique. Additionally, SILRAD consumed considerably less memory and operated faster than these other methods, even when faced with an imbalanced dataset where ransomware constituted the minority class. This work paves the way for future research to create a real-time ransomware detection system, emphasising the importance of adaptive learning in combating sophisticated cyber threats.
\end{abstract}




\maketitle
\section{Introduction}

Since the beginning of the COVID-19 crisis, malware has become a global challenge \cite{ispahany2021detecting}. One particular strain of malware that has gained notoriety recently is ransomware, which encrypts sensitive files to sell the decryption keys back to the victim. Despite improvements in malware detection, ransomware attacks have only increased. During the first half of 2021, ransomware payments totalled 590 million USD in the United States alone, exceeding the \$416 million worth of losses suffered the year before \cite{treasury_2021}.\\
The first known ransomware attack occurred in 1989 with the emergence of the AIDS ransomware. This ransomware encrypted victims' files and demanded a payment of \$189, purportedly for AIDS research, in exchange for the decryption keys. This attack laid the foundation for developing more sophisticated ransomware variants in the following years \cite{al2018ransomware}. Since then, ransomware has become a more sophisticated threat and a lucrative business for organised criminal groups. \\
To address the growing ransomware threat, researchers have investigated various detection techniques to identify ransomware activity before sensitive data is encrypted. These techniques include signature-based methods, which analyse static data extracted from systems or files at rest, and dynamic analysis approaches, which allow suspicious files to execute organically to monitor system or file activity for ransomware behaviour. Recent studies have increasingly focused on dynamic analysis due to the ability of modern ransomware to evade signature-based detection methods. More recently, machine learning (ML) has been employed to automate this process by training ML models on known ransomware samples to classify new data as either malicious or benign \cite{urooj2021ransomware}. Researchers have proposed a variety of machine learning techniques for ransomware detection, ranging from traditional algorithms such as Support Vector Machines (SVM) to more advanced deep learning approaches like Long Short-Term Memory (LSTM) networks \cite{ispahany2024ransomware}.
However, despite extensive research efforts, most proposals in the literature have utilised ML in a non-incremental setting, requiring the underlying model to be completely rebuilt from scratch to incorporate new ransomware strains and remain effective. \\
While non-incremental approaches have demonstrated impressive detection accuracies, the rapid proliferation of ransomware strains makes retraining the model with old and new data for each update time-consuming and resource-intensive. Importantly, non-incremental ML techniques may not be able to detect the adaptations emerging over time as ransomware evolves to evade detection, leaving sensitive data vulnerable to attack. In other words, relying on outdated ML models for ransomware detection poses significant risks to sensitive data. Moreover, most studies focus on detecting ransomware in bulk datasets, often after the ransomware has been executed, rather than in real-time data streams. This delay creates a critical window in which ransomware can operate undetected. The presence of a substantial "training gap," the inability to adapt to evolving ransomware strains, and the lack of real-time detection capabilities in traditional non-incremental ML approaches highlight their limitations and make them impractical for effective ransomware detection.\\ 
Unlike static ML techniques, online incremental learning can continuously learn new knowledge from new data whilst maintaining most of the previously learnt knowledge without having to rebuild the models from scratch, provided they have suitable concept drift detection.  This feature ensures that the underlying ML model is continuously updated to defend against newly discovered ransomware attacks. These characteristics prompted the authors to develop a ransomware detection system based on online incremental learning, which can quickly adapt to evolving ransomware without requiring the entire ML model to be retrained from scratch. \\
In light of this, the contributions of this paper are as follows: \\
\begin{itemize}
    \item \textit{Incremental learning}: We propose a novel technique to detect evolving ransomware using online incremental learning called Sysmon Incremental Learning System for Ransomware Analysis and Detection (SILRAD). We show how our method is superior to traditional ML techniques, which cannot detect new ransomware unless the model is trained from scratch
    \item \textit{Useful features}: We identify the most critical features within Sysmon logs to detect contemporary ransomware behaviour accurately
    \item \textit{Lab setup and ransomware dataset}: We outline the lab setup we used to safely extract Sysmon logs from live ransomware harvested from the wild to aid future researchers
    \item \textit{Algorithm comparison}: We compare SILRADs detection accuracy with traditional ML techniques proposed throughout the literature and show its effectiveness in detecting new ransomware without updating the model from scratch. We also evaluate incremental algorithms considered for SILRAD and compare their detection accuracy, memory consumption and classification times 
\end{itemize}

The rest of this paper is organised as follows: Section \ref{related-work} highlights related work on ransomware, its detection using incremental learning and Sysmon, Section \ref{approach} outlines the proposed technique, Section \ref{sec:experiment} discusses the experiment in detail, and Section \ref{sec:results} presents the results and show that SILRAD is superior to traditional ML approaches at detecting new ransomware.

\section{Related work}\label{related-work}


\subsection{Ransomware behaviour and detection methods}

Ransomware is a form of malware designed to extort victims by encrypting sensitive files and demanding payment for the decryption key. It primarily exists in two forms: locker ransomware, which restricts access to the system, and crypto-ransomware, the more prevalent type that utilises cryptography to render files inaccessible. The attack lifecycle consists of several stages, beginning with reconnaissance, where attackers gather information about the target system. The ransomware is then delivered, typically through phishing emails or malicious websites, and installed by exploiting system vulnerabilities. Following installation, the ransomware often connects to an external command and control (C\&C) centre, allowing the attackers to communicate with the compromised system. In the final stage, the ransomware encrypts the targeted files and demands a ransom for decryption \cite{barnum2012standardizing}.\\
Modern variants have evolved to adopt more sophisticated strategies, such as double extortion, where attackers encrypt data and threaten to leak sensitive information if the ransom is not paid \cite{rhysida-cisa}. Contemporary ransomware also commonly employs advanced evasion techniques to bypass detection systems, such as obfuscation, DLL-sideloading, and running directly in memory via scripts \cite{d2021rope, chakkaravarthy2019survey, afianian2019malware, kumar2020emerging, avoslocker_ransomware_2022}. It also utilises symmetric and asymmetric encryption methods to enhance the effectiveness of its attacks \cite{bajpai2020attacking}.\\
Throughout the literature, researchers have proposed various techniques for detecting ransomware, typically occurring during its delivery to the target system \cite{liu2022netsentry, berrueta2022crypto, fernandez2019intelligent}, execution on the system \cite{homayoun2019drthis, shaukat2018ransomwall, roy2021deepran}, or communication with external systems for exfiltrating sensitive data \cite{almashhadani2020maldomdetector}. Early detection methods relied on static analysis techniques, such as extracting file attributes like MD5 hashes, Opcodes, entropy, or printable strings from files at rest \cite{ye2010automatic, santos2011semi} and comparing them to known malicious files. Although straightforward, this approach became less effective as ransomware evolved and began using obfuscation techniques to evade detection, making signature-based methods increasingly challenging \cite{singh2020survey}. As a result, the focus shifted to behaviour-based detection using dynamic analysis techniques. Unlike static analysis, dynamic analysis monitors system activity during execution in a controlled environment to detect suspicious behaviour associated with ransomware \cite{ispahany2024ransomware}. Machine learning has recently become essential in automating this process by training models on known ransomware behaviour to identify new threats and classify system or file data as either malicious or benign.

\subsection{Ransomware detection using online incremental learning}

Machine learning (ML) has significantly contributed to automating ransomware detection by enabling researchers to train models on known ransomware behaviours and identify emerging threats. Various ML techniques have been proposed in the literature, ranging from traditional algorithms to more advanced deep-learning approaches. Traditional ML methods, such as Support Vector Machines (SVM) \cite{hsu2021enhancing}, Decision Trees (DT) \cite{almashhadani2020maldomdetector, ahmed2020system, berrueta2022crypto}, and Random Forests (RF) \cite{kok2020early, khammas2020ransomware, kok2020evaluation}, have been widely adopted due to their minimal resource requirements, good accuracy, and ease of implementation. More recent research has focused on advanced deep learning techniques like Convolutional Neural Networks (CNN) \cite{chaganti2023multi, gulmez2024xran, ciaramella2023explainable, cen2024zero}, Recurrent Neural Networks (RNN) \cite{li2022machine, rhode2021real}, and Long Short-Term Memory Networks (LSTM) \cite{homayoun2019drthis, molina2021ransomware, woralert2023hard, davidian2024early}, which have demonstrated higher accuracy and potential for real-time detection, albeit with greater resource demands. \\
Despite the promising results of these methods, most rely on offline batch learning techniques that require the model to be retrained from scratch to incorporate new ransomware strains, which limits their adaptability in rapidly evolving threat landscapes \cite{darem2021adaptive}. This is problematic since the volume of active ransomware on the internet is constantly growing, and static ML techniques without concept-drift detection are unaware when the underlying ML model needs repairing or building. As a result, sensitive data is vulnerable to attack until the model is updated. \\ 
An alternative approach is to update the ML model incrementally, thus removing the dependency on retraining the model from scratch. Unlike static ML techniques, incremental learning models are often equipped with suitable concept drift detection, giving them awareness when old models require repairing or rebuilding. This makes it the preferred approach for ransomware detection. Despite this, detecting ransomware activity using incremental learning amongst evolving data streams is an understudied domain. 
Several authors throughout the literature have proposed detection systems based on batch-based incremental learning \cite{darem2021adaptive, li2020incremental, roy2021deepran}. However, none of these studies can detect ransomware amongst online data streams.\\
Li \textit{et al.} \cite{li2020incremental} propose a ransomware detection system based on a multi-class SVM that is updated in batches of 50 malware samples. However, the model requires more than 1,000 malware samples from each class to obtain the required accuracy. This makes the system ineffective in detecting newly discovered malware or ransomware specifically designed for a particular attack and has less than 1,000 samples in the wild.   \\
Several deep-learning-based incremental learning ransomware detection systems have also been proposed throughout the literature \cite{darem2021adaptive, roy2021deepran}. However, deep-learning approaches are typically resource-intensive and require long training times \cite{yang2022adaptability}. This can be a risk for sensitive data if the training time is prolonged, especially if a ransomware attack occurs within the "training gap". \\
Darem \textit{et al.} \cite{darem2021adaptive} propose an Adaptive Behavioral-Based Incremental Batch Learning Model for detecting malware variants to address the challenges posed by concept drift resulting from evolving malware. The model, grounded in sequential deep learning, analyses API sequences to identify malware activity and is updated when detecting concept drift. The authors report a high detection accuracy of 99.41\% and a low updating frequency of 1.35 times per month. Despite this, the batch-based update mechanism limits its ability to detect ransomware in real-time data streams.\\
Roy and Chen \cite{roy2021deepran} use a bidirectional long short-term memory network (BiLSTM) model and update it in mini-batches of 128 events. However, like the studies above, it can also not handle online data streams. \\
As can be seen from the examples mentioned earlier, online incremental learning to detect ransomware activity within data streams has seldom been studied in the literature. This presents an opportunity to explore its utility in detecting malicious behaviour amongst data streams to mitigate the risk of a ransomware attack while maintaining a low resource footprint. 

\subsection{Leveraging Sysmon logs for ransomware detection}\label{sysmon} 

The System Monitor (Sysmon) is a security utility incorporated in the Sysinternals package for Windows operating systems (O/S) \footnote{ https://learn.microsoft.com/en-us/sysinternals/downloads/sysmon}. Operating as a Windows system service, the Sysmon device driver dispatches security events, such as process tampering and file terminations, to the Windows event log. Sandboxes and virtual machines have frequently been used throughout the literature to harvest features such as API calls and Opcodes post-detonation to aid classification. However, advanced malware is often equipped to obfuscate malicious behaviour and even detect the analysis environment, making ransomware analysis less effective \cite{caviglione2020tight}. 
Sysmon, on the other hand, provides real-time security-related events that are difficult to obfuscate since it operates on the executive level of the operating system. Despite its widespread use within enterprise environments for threat detection and analysis \cite{mavroeidis2018data}, few studies have used ML-based approaches to detect ransomware activity within Sysmon logs \cite{smiliotopoulos2022revisiting}. \\
Grimshaw \textit{et al}. \cite{grimshaw2024link} utilised Sysmon to detect suspicious behaviours indicative of security threats, such as unauthorised access or malicious activities. Their study specifically focused on Sysmon ID 3, which provides detailed information about network behaviour. The authors developed a Graph Attention Network-Graph Autoencoder (GAN-GAE) model and ran attack scenarios such as Advanced Persistent Threats (APT-29) across four corporate networks. \\
Xuon and Huong \cite{do2022new} also presented a deep graph network-powered approach to detect Advanced Persistent Threat (APT) activity within Sysmon logs. 
The authors use Sysmon to detect Advanced Persistent Threats (APTs) to address the limitations of sandbox environments in identifying ransomware. Traditional sandboxes may fail to detect the stealthy evasion techniques often used by APTs. By contrast, Sysmon allows for a more direct analysis of system behaviours, potentially uncovering hidden malicious activities that sandboxes might overlook.
The authors use various algorithms, such as Graph Convolutional Networks (GCN) and Graph Isomorphism Networks (GIN), to model and classify malicious behaviour with an accuracy exceeding 90\%.\\
Other studies have used Sysmon to accurately detect malicious behaviour without using ML-based approaches. Mavroeidis and J{\o}sang \cite{mavroeidis2018data} propose a non-ML-based approach to detect ransomware activity amongst a continuous stream of Sysmon logs. In their study, the authors automatically assign threat levels to Sysmon events. The proposed system queries the knowledge base of ransomware activity for indicators of compromise (IOCs) using SPARQL. \\
Although the studies mentioned above have not specifically utilised Sysmon for ransomware detection, they have demonstrated Sysmon's capability to provide a reliable stream of real-time events within the Windows operating system that can be used to classify malicious behaviour. This motivates the authors to leverage Sysmon for ransomware detection, which is the central focus of this study.

\section{Proposed approach}\label{approach}
\begin{figure*}[t!]
    \centering
    \includegraphics[width=0.95\linewidth]{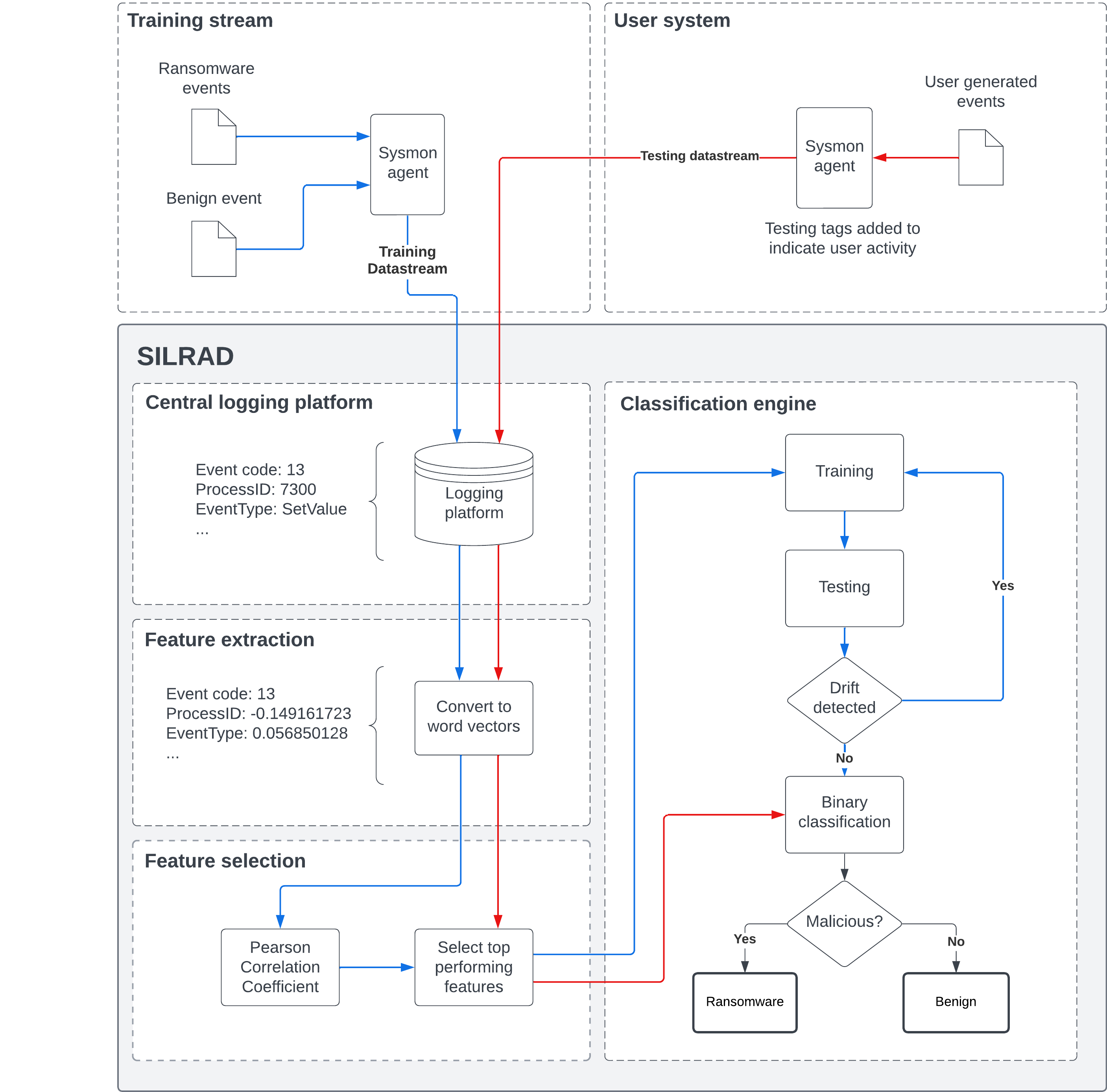}
    \caption{The Proposed Sysmon Incremental Learning System for Ransomware Analysis and Detection (SILRAD). The blue arrows indicate the data stream used to train the model, and the red arrows indicate system activity to be classified }
    \label{fig:SILRAD}
\end{figure*}

This section discusses the Sysmon Incremental Learning system for Ransomware Analysis and Detection (SILRAD) technique as presented by Fig. \ref{fig:SILRAD}, a novel approach to detecting ransomware activity within streams of Sysmon logs using online incremental learning as explained in Algorithm. \ref{alg:cap}. Our approach utilises behaviour-based detection techniques via dynamic analysis to analyse events generated by Sysmon from Windows system activity. This method addresses the limitations of existing approaches as outlined in Section \ref{related-work}. These events are sent to a centralised logging server and automatically classified as benign or malicious using online incremental learning. Unlike traditional methods, our approach continuously updates the machine learning model whilst handling concept drift, allowing for the detection of new ransomware strains without the need for complete model retraining. This ensures adaptability and efficiency in real-time threat detection.

\begin{algorithm*}
\caption{SILRAD \textbf{Symbols:} $E$: The Sysmon event, $\textbf{x}$: input feature values, $y$: target class label, $W$: Window of values where $w_0$ is the previous window and $w_1$ is a current window, $\epsilon_{cut}$: The drift threshold value }\label{alg:cap}
\begin{algorithmic}

\Function{}{}
    \State Initialise Window W
    \While{HasNext($E$)}
        \State $(\textbf{x}, y) \gets$ next$(E)$
        \For{All $x, y \in E$}
            \State $(\hat{\textbf{x}}) \gets $ ConvertSentanceToVector($\textbf{x}$)
            \State CalculatePCC$(\hat{\textbf{x}}, y)$ \Comment{Calculate the Pearson Correlation Coefficient}
            \State SelectTopFeatures$(\hat{x}, y)$
            \State $\Ddot{y} \gets $Predict($\hat{x}$) \Comment{Predict the target class}
            \If{$\Ddot{y} =$ ransomware}
                \State RansomwareClass
            \EndIf
            \State UpdateModel($\hat{x}, y$)
            \State $W \gets W \cup \{\hat{x}\}$ \Comment{adds $\hat{x}$ to the end of $W$}
            \Repeat 
                \State Drop elements from end of $W$
            \Until{$\vert\hat{\mu}_W{_0} - \hat{\mu}_W{_1}\vert \geq \epsilon_{cut}$} \Comment{i.e Cuts the tail if the average values are $\geq$ threshold}
            
        \EndFor
    \EndWhile
    
\EndFunction
\end{algorithmic}
\end{algorithm*}

\newpage
\subsection{Dynamic analysis using Sysmon}
Employing dynamic analysis mitigates the vulnerabilities of static analysis, such as obfuscation and polymorphism. Subsequently, researchers have explored a broad spectrum of dynamic features to ascertain ransomware behaviour. Native Windows events, such as API calls and Opcodes, have successfully been used throughout the literature to divulge ransomware behaviour dynamically, albeit with challenges mentioned in section \ref{sysmon}. Conversely, Sysmon operates at the executive level of the O/S (making techniques such as API spoofing more cumbersome) and is scalable, reliable, easy to install, and resilient to reboots. Importantly, Sysmon provides an avenue to extract system logs to a central logging platform without using well-known analysis tools that are easily detected by contemporary malware.
For this reason, we leverage Sysmon to supply a steady stream of security events to classify ransomware behaviour. Although custom rules can be configured, we propose using the default events to measure the efficacy of Sysmon events as a feature. By default, Sysmon offers 29 inbuilt events to indicate suspicious behaviour, some of which can be seen in Table. \ref{tab:table-sysmon}. Any events not triggered throughout the experiment were redacted from the table to improve relevance. Importantly, although Sysmon discloses security-related events, not all are malicious and legitimate applications may trigger Sysmon events. 

\begin{table}[]
\centering
\caption{A list of Sysmon events and their descriptions observed throughout the experiment}
\label{tab:table-sysmon}
\resizebox{0.6\linewidth}{!}{%
\begin{tabular}{|c|l|l|}
\hline
\textbf{Event} &
  \textbf{Event name} &
  \textbf{Description} \\ \hline
1 &
  Process creation &
  Full details about a newly created process \\ \hline
2 &
  \begin{tabular}[c]{@{}l@{}}A process changed a file \\ creation time\end{tabular} &
  \begin{tabular}[c]{@{}l@{}}Triggered when a file creation time is \\ modified by a process\end{tabular} \\ \hline
3 &
  Network connection &
  TCP/UDP connections on the machine \\ \hline
5 &
  Process terminated &
  \begin{tabular}[c]{@{}l@{}}Triggered when a process is terminated \\ on the machine\end{tabular} \\ \hline
7 &
  Image loaded &
  \begin{tabular}[c]{@{}l@{}}Triggered when a module (dd) is loaded\\ in a specific process\end{tabular} \\ \hline
8 &
  CreateRemoteThread &
  \begin{tabular}[c]{@{}l@{}}Triggered when hiding techniques are \\ detected such as process hollowing\end{tabular} \\ \hline
10 &
  ProcessAccess &
  \begin{tabular}[c]{@{}l@{}}Reported when a process opens another \\ process\end{tabular} \\ \hline
11 &
  FileCreate &
  \begin{tabular}[c]{@{}l@{}}Logged when a file is created on the \\ system or overwritten\end{tabular} \\ \hline
12 &
  \begin{tabular}[c]{@{}l@{}}RegistryEvent (Object \\ create and delete)\end{tabular} &
  \begin{tabular}[c]{@{}l@{}}Triggered when a registry key is created \\ or deleted\end{tabular} \\ \hline
13 &
  RegistryEvent (Value set) &
  \begin{tabular}[c]{@{}l@{}}Logged when values in the systems \\ registry are modified\end{tabular} \\ \hline
17 &
  PipeEvent (Pipe created) &
  Generated when a named pipe is created \\ \hline
22 &
  DNSEvent (DNS query) &
  \begin{tabular}[c]{@{}l@{}}Reported when a process executes a \\ DNS query\end{tabular} \\ \hline
23 &
  \begin{tabular}[c]{@{}l@{}}FileDelete (file delete \\ archived)\end{tabular} &
  Logged when a file is deleted \\ \hline
25 &
  \begin{tabular}[c]{@{}l@{}}ProcessTampering \\ (process image change)\end{tabular} &
  \begin{tabular}[c]{@{}l@{}}Triggered when hiding techniques are \\ detected such as process hollowing\end{tabular} \\ \hline
\end{tabular}%
}
\end{table}

\subsection{Feature representation using fastText}
\label{fasttext}
\begin{figure}
    \centering
    \includegraphics[width=.4\linewidth]{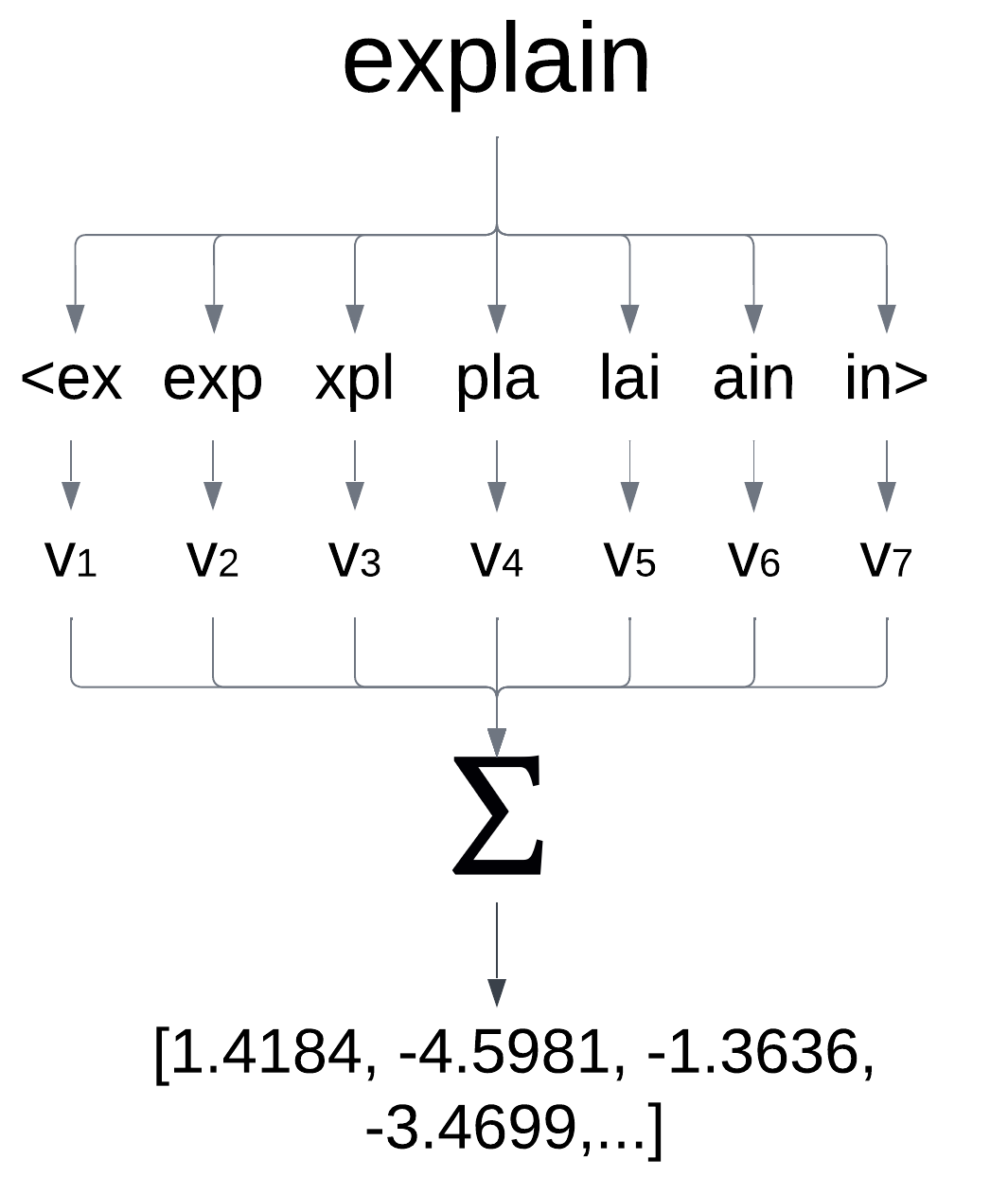}
    \caption{fastText conversion to vectors using n-grams. The above example shows the process to convert the word "explain" into vectors where $v_n$ represents the vector representation of the corresponding $n$-gram}
    \label{fig:fasttext}
\end{figure}

We use fastText to convert words into vectors for the feature representation of the model. Other embedding techniques, such as Word2vec, represent words as a whole and ignore their internal structure, a limiting factor for rare words. On the other hand, fastText breaks words into character $n$-grams as shown in Fig. \ref{fig:fasttext}, and represents words as the vector sum of these $n$-grams, improving the accuracy for more complex words \cite{bojanowski2017enriching}. 

Consider a set of $n$-grams within a word represented by $\mathcal{G}_{w} \subset \{1, \cdots, G\}$ where $G$ represents the length of the dictionary set. Each corresponding $n$-gram, $g$, is associated with a vector $\textbf{z}_g$, with the dimensions of $N \times 1$. Let us represent the surrounding context words by a vector $\textbf{v}_c$ with dimensions of $N \times 1$ as well. Now, we can define the scoring function as 

\begin{align}
  s(w, c) =  \sum_{g \in \mathcal{G}_{w}}^{|G|} \bigg ( \textbf{z}^T_g\textbf{v}_c \bigg ) = 1 
  \label{eqn:SF}
\end{align}

This is useful when converting unique values within the Sysmon logs, such as path name, which other techniques like Word2vec may not recognise. By default, Fasttext creates embeddings with a vector size of 100. However, to simplify and reduce the preprocessing required, we calculated the average of word vectors with a dimension size of 100. 

\subsection{Feature selection technique}\label{selection}
Since the goal of SILRAD is to detect ransomware activity as it arrives from Sysmon, an essential factor in the design of SILRAD is to reduce unnecessary preprocessing of features. By default, Sysmon logs consist of 52 features, so to achieve this, we leverage the Pearson Correlation Coefficient (PCC) intermittently to measure the similarity between the input feature values and the target class values and select the most significant features. This allows SILRAD to dynamically update the most critical feature sets in response to changes in the data. It's important to note that the actual value of the target class is necessary to apply this equation. Therefore, we employ this mechanism exclusively within the training stream, as illustrated in the Fig. \ref{fig:SILRAD}. Mathematically, this is calculated as the normalisation form of the covariance of two variables $x$ and $y$ where $x$ represents an input feature value and $y$, which represents the equivalent target class value or true value:

\begin{align}
     PCC(x,y) = \frac{cov(x,y)}{\sigma_x \sigma_y} = \frac{E[(x - \overline{x})(y - \overline{y})]}{\sigma_x \sigma_y}
\end{align}

\begin{figure}
    \centering
    \includegraphics[width=0.8\linewidth]{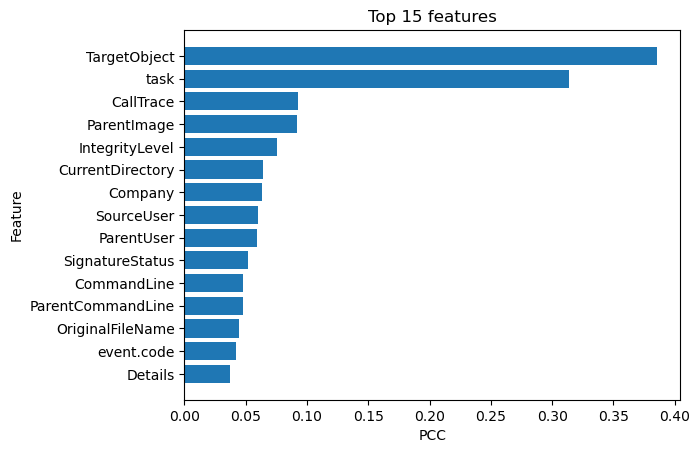}
    \caption{The most significant Sysmon features calculated by the Pearson Correlation Coefficient (PCC)}
    \label{fig:features}
\end{figure}

Where $\overline{x}$ and $\overline{y}$ are the averages of $x$ and $y$; $\sigma_x$ and $\sigma_y$ are the standard deviations, respectively \cite{Wang2013}. As seen in Fig. \ref{fig:features}, overall, the most significant features were the TargetObject, Task, CallTrace and ParentImage. A potential limitation of our model is that key features may change over time, and the model may not initially include historical data for new classes. This can lead to reduced classification accuracy until the model is retrained. We aim to tackle this challenge in our future research.

\subsection{Online incremental learning}\label{incremental}

Traditional offline machine learning operates under the premise that the data is provided before training. This means the underlying model has the luxury of being trained on the full dataset, assuming the data is static. By contrast, incremental learning must adapt to a constantly arriving dataset and learn new knowledge from new samples whilst maintaining previously learned knowledge. Since ransomware constantly evolves, the underlying ML model must constantly be updated to detect new ransomware adequately and effectively. With this backdrop, we leverage online incremental learning to classify inbound Sysmon data streams. Conceptually, the objective is to predict a class label $\hat{y}_t$ for a very large sequence, $S = \{s_1, s_2, ..., s_T\}$ where $T$ has a large value, of arriving tuples $s_i = (\textbf{x}_t, y_t)$. The prediction, $\hat{y}_t$, is determined using its corresponding input value of $\textbf{x}_t$ with the previously learned model $m_{t-1}$, where $\textbf{x}_t$ represents an instance in the $n \times 1$ dimensional feature space, characterised by $\textbf{x}_t = [x_1, x_2, ..., x_n]$ at the current timestamp $t$. This is represented by Equation \ref{eqn:model}:

\begin{align}
    \hat{y}_t = \sum_{t > 1}^{|S|} \bigg( m_{(t-1)} \Rightarrow (\textbf{x}_t) \bigg) ,
    \label{eqn:model}
\end{align}

where $|S|$ represents the cardinality of $S$ and $\Rightarrow$ represents the learning pocess of model $m_{(t-1)}$ through the input values of $\textbf{x}_t$. Also, it is important to note that the start point of the learning is at $t > 1$ since the initial learned model at $t \le 1$ does not produce accurate predictions for $\hat{y}_0$, as can be seen in the results in Fig. \ref{fig:mcc-graph}. For this reason, it is assumed real-world scenario, classification begins at time $t > 1$.

\begin{figure}
    \centering
    \includegraphics[width=0.6\linewidth]{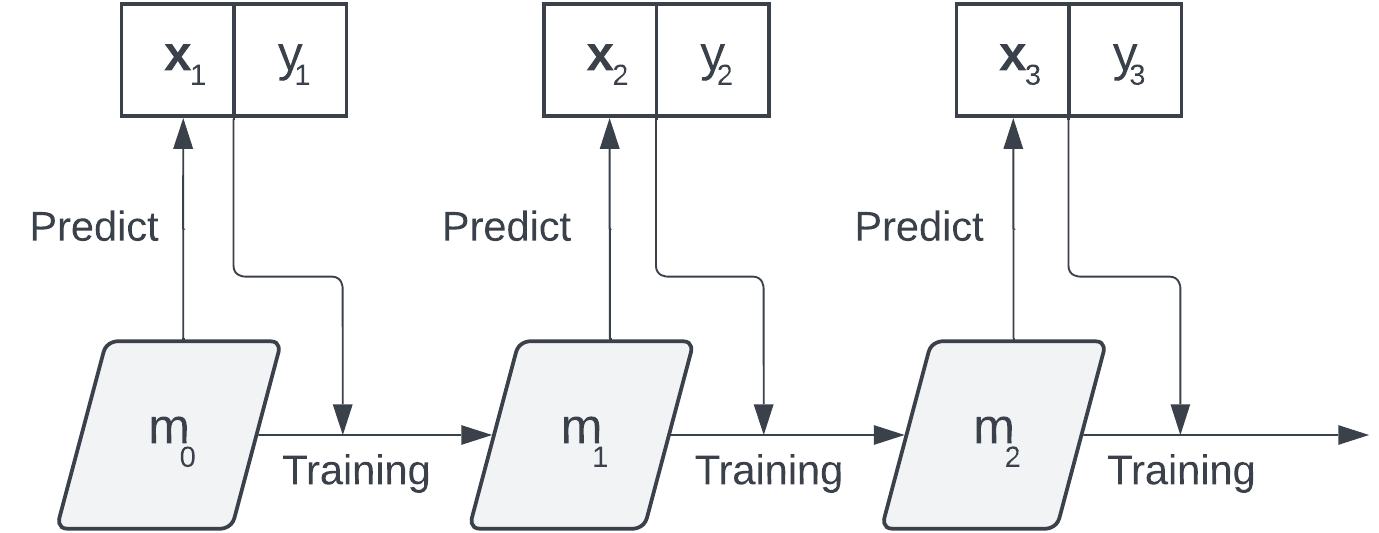}
    \caption{The online learning model used by SILRAD whereby the model is trained and predictions made per instance of data arriving}
    \label{fig:online-model}
\end{figure}

The above equation, coupled with suitable concept drift detection, allows SILRAD to quickly make class label predictions whilst adapting to new data without retraining the model from scratch, as shown in Fig. \ref{fig:online-model}. This is an important requirement to prevent unauthorised encryption of sensitive data from new ransomware, which otherwise may occur if the model is trained in batches. An additional benefit is that fewer resources are consumed since incremental learning models do not need to store the training set in memory, unlike many batch models \cite{bernardo2020incremental}. \\
In line with the above, we select the ensemble method, Adaptive Random Forest (ARF), for the base classifier for SILRADs classification engine due to its ability to process evolving data streams, good classification performance and concept drift detection \cite{gomes2017adaptive}.  

\subsection{Concept drift detection}
A key component of SILRAD is automatically updating the model when degradation is detected. Concept drift detection is required to reset the underlying model when a drift signal is triggered. To facilitate this, we employ the Adaptive Windowing (ADWIN) technique to monitor for changes in the data stream. ADWIN is a popular concept drift detection technique using sliding windows that dynamically grow when no changes are detected and shrinks when differences are present \cite{bifet2007learning}. The primary mechanisms underpinning ADWIN include window adjustment, cut detection, and error bounding. The window adjustment process involves retaining recently observed data within a variable-sized window, thereby facilitating continuous monitoring of data fluctuations. Cut detection is a critical element that determines the appropriate moments to contract the window; this occurs when there is a significant difference in the statistical properties of the data segments before and after a specific point. Finally, the error bound component is crucial in deciding whether the differences between two compared windows are substantial enough to warrant a cut. This methodology enables ADWIN to effectively track and adapt to changes, making it an essential tool for real-time data analysis applications. Although one disadvantage of ADWIN is that it requires access to labelled data \cite{gomes2019streaming}, we accept its usage for SILRAD because Sysmon data is accurately labelled once ingested.
\\
To illustrate how ADWIN detects changes in the data stream, consider a sliding window $W$ of length $n$, with two sub-windows $W_0$ and $W_1$ of lengths $n_0$ and $n_1$, respectively. The harmonic mean $m$ of $n_0$ and $n_1$ is firstly calculated using the following formula:

\begin{align}
    m = \frac{2}{1/n_0 + 1/n_1}
    \label{eqn:mean}
\end{align}

The error bound $\epsilon_{cut}$ value is determined using a desired confidence level $\delta$ and the following formula:

\begin{align}
    \epsilon_{cut} = \sqrt{\frac{1}{2m}\cdot\ln\frac{4}{\delta}}
    \label{eqn:ecut}
\end{align}

Finally, we compare the absolute differences between the means of the values in the sub-windows $W_0$ and $W_1$ against the calculated error bound. If this difference is statistically significant, $W_0$ is discarded, retaining only $W_1$ and any new data that arrives. For example, consider a data stream $W$ containing the following values:

\begin{align}
    W = [1,2,2,2,3,4,5,5,6,7,8,9]
    \label{eqn:array_w}
\end{align}

The window is divided into two separate sub-windows $W_0$ and $W_1$. In this example, we divide these windows equally:

\begin{align}
    W_0 = [1,2,2,2,3,4] \\
    W_1 = [5,5,6,7,8,9]
    \label{eqn:array_w}
\end{align}

The harmonic mean is then calculated as follows:

\begin{align}
    m = \frac{2}{1/n_0 + 1/n_1} = \frac{2}{1/6 + 1/6} = 6
    \label{eqn:mean}
\end{align}

Assuming the confidence level $\delta$ = 0.05, the error-bound calculation is determined as per the following formula:

\begin{align}
    \epsilon_{cut} = \sqrt{\frac{1}{2m}\cdot\ln\frac{4}{\delta}} = \sqrt{\frac{1}{2.6}\cdot\ln\frac{4}{0.05}} \approx \sqrt{\frac{1}{12}\cdot5.99} \approx \sqrt{0.499} \approx 0.706
    \label{eqn:ecut}
\end{align}

Next, the means of the values of each sub-window are calculated:

\begin{equation}\label{eqn:means}
    \begin{aligned}
        mean(W_0) = \frac{1 + 2 + 2 + 2 + 3 + 4}{6} = \frac{14}{6} \approx 2.33 \\
        mean(W_1) = \frac{5 + 5 + 6 + 7 + 8 + 9}{6} = \frac{40}{6} \approx 6.67
    \end{aligned}
\end{equation}

Finally, the absolute difference between the averages are compared with the error-bound calculation:

\begin{align}
    \vert mean(W_0) - mean(W_1) \vert = \vert 2.33 - 6.67 \vert = 4.34
    \label{eqn:absolute}
\end{align}

Since the absolute difference between the averages exceeds the error-bound value (4.34 > 0.706), $W_0$ is removed from $W$, making room for new data. This adjustment allows the window to adapt dynamically to significant changes in the data stream.

\section{Experiment}\label{sec:experiment}

\subsection{Experiment setup}

\begin{figure*}[t!]
    \centering
    \includegraphics[width=.85\linewidth]{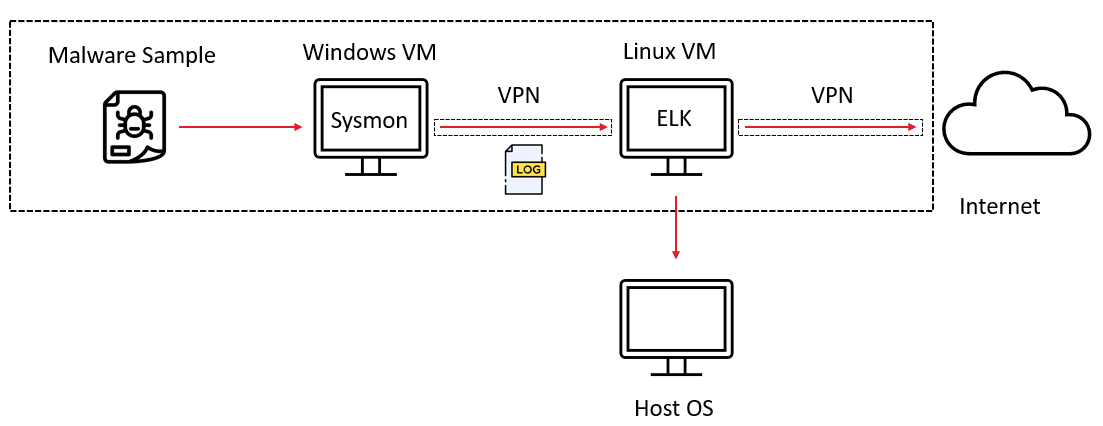}
    \caption{The experiment setup used to detonate ransomware and harvest features}
    \label{fig:lab}
\end{figure*}

A common approach to analyse ransomware behaviour throughout the literature is detonating samples on sandboxes such as Cuckoo to harvest logs post-execution. The challenge of this approach is that contemporary ransomware is often aware of its operating environment and restrains execution to thwart detection. Accordingly, we designed our lab environment to entice ransomware to execute organically by replicating a production endpoint with full internet access, as shown in Fig. \ref{fig:lab}. We deploy isolated virtual machines (VM) with Windows 11 to facilitate this. Although the preferred approach is to detonate ransomware on bare metal, virtual machines have the flexibility to be easily rolled back to a previous restore point post-infection. \\
To simulate a production system, over 1000 dummy files were copied to the file system from filesamples \footnote{https://filesamples.com/} consisting of csv, gif, jpeg, mp4, png, ppt, pdf, docx and zip files.  
\\
Sysmon agents were deployed on Windows virtual machines on the network to forward Sysmon events to SILRAD as explained in section \ref{SILRAD}.

\subsection{SILRAD environment}\label{SILRAD}
The core infrastructure of SILRAD is hosted on an Ubuntu Server 22.04 LTS, equipped with 4 CPUs, 12GB of RAM, and 80GB of disk space. SILRAD operates through two primary components, as illustrated in Fig. \ref{fig:SILRAD}: a central logging server and an incremental learning platform. The incremental learning platform comprises of a feature extraction module, a feature selection component, and a classification engine. The central logging server uses the Elasticsearch, Logstash, and Kibana (ELK) stack, where Sysmon events from individual agents are ingested for processing. These events undergo fastText preprocessing before being streamed to the feature extraction module and, subsequently, the classification engine. Despite the Sysmon events being sent remotely to the ELK stack, the time delay between the event occurrence and its appearance on the logging server was under 1 second.\\
The classification engine utilises River 0.19.0 \cite{montiel2021river} for applying online incremental learning and is underpinned by Python 3.9.13.

\subsection{Network considerations}\label{network}
Ransomware typically communicates with external command and control centres (C\&C) to download instructions, transfer encryption keys and exfiltrate data. However, some modern ransomware won't entirely execute if an analysis environment is detected or it cannot contact external C\&C's. To ensure the ransomware executes completely during the experiment, communication lines to external C\&Cs must be open whilst preventing the ransomware from spreading to other nodes. Facilitating this communication requires careful consideration to prevent further attacks from external attackers on the internal lab environment. For example, detonating ransomware samples from a legitimate IP address within the network could pave the way for future attacks on legitimate nodes.
For this reason, network traffic from the VM was channelled through a Linux gateway and connected to the internet via a virtual private network (VPN), as shown in Fig. \ref{fig:lab}. This approach also can change the source IP address if compromised. Another factor considered when building the ransomware lab was preventing infection across the network. Even though some recent ransomware has been known to infect Linux systems, only ransomware known to impact Windows systems was selected to avoid cross-contamination.

\subsection{Dataset}
\label{subsec:dataset}

\begin{figure}[htbp]
  \centering
  \begin{tabular}{@{}c@{}}
    \includegraphics[width=0.65\linewidth]{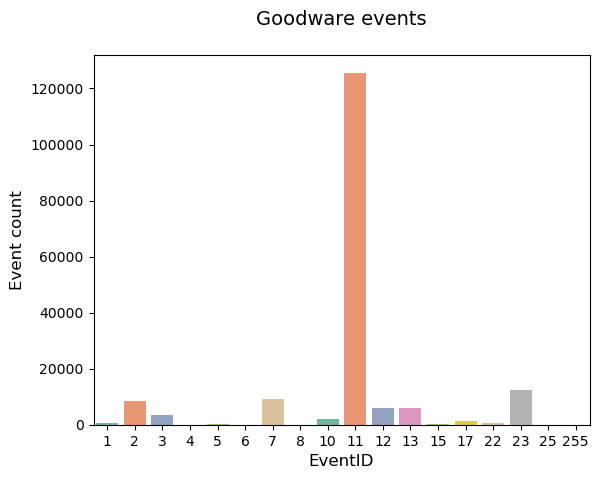} \\[\abovecaptionskip]
    \small (a) Sysmon events triggered from the execution \\ of goodware samples
  \end{tabular}

  \vspace{\floatsep}

  \begin{tabular}{@{}c@{}}
    \includegraphics[width=0.65\linewidth]{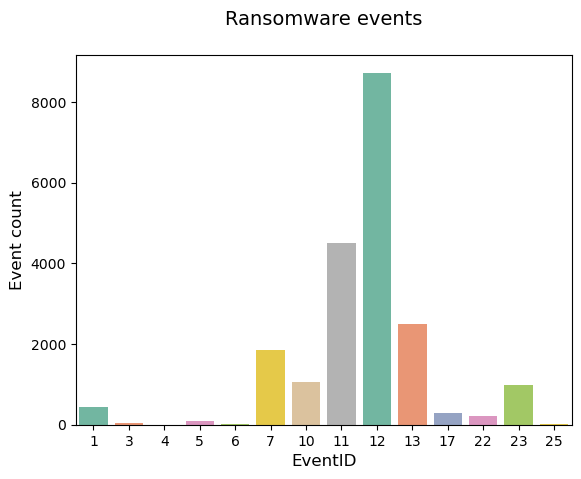} \\[\abovecaptionskip]
    \small (b) Sysmon events triggered from the execution \\ of ransomware samples
  \end{tabular}

  \caption{Number of Sysmon events triggered from both goodware and ransomware by eventID }\label{fig:events}
\end{figure}

An imbalanced dataset was constructed to evaluate SILRAD's capability to detect new ransomware without requiring complete model retraining. Given the absence of existing Sysmon ransomware datasets in the literature, the dataset was created using a combination of goodware and ransomware samples from six well-known ransomware families. Doing this required capturing Sysmon events from the start of its execution until the ransomware note is displayed. However, creating a continuous stream of benign and ransomware events is challenging. Each ransomware sample renders the virtual machine unusable post-execution, requiring it to be rolled back to a previous restore point. For this reason, Sysmon events generated from goodware and ransomware were "stitched" together to train and test the models. \\
In line with recent attacks, modern ransomware samples were collected from VirusTotal and HybridAnalysis, such as AvosLocker, BlackBasta, Conti, Hive, Lockbit, and REvil. These samples were chosen due to their destructive nature and the volume of attacks recently observed in the wild using these ransomware families. In total, over 20,000 Sysmon events were harvested from 50 ransomware attacks. \\
Each ransomware sample was executed for 5 minutes to maintain a consistent approach. However, in most cases, the ransomware note was displayed before this time. Once executed, Sysmon logs were automatically forwarded to a central logging server until the virtual machine was locked and displayed the ransomware note, which, in most cases, prevented further events from being logged.\\
Typically, benign events outweigh malicious events in real-world attacks. With this presumption, we created an imbalanced dataset by continuously harvesting benign events from various games, utilities and applications collected from PortableApps. This process continued until a corpus of over 176,000 benign events were harvested. \\
The final dataset contained nearly 200,000 ransomware and benign events as shown in Table. \ref{tab:totalevents}, with blocks of ransomware activity positioned throughout. We simulate drift by introducing new ransomware families throughout the dataset to measure SILRAD's ability to adapt to evolving ransomware. An overview of the Sysmon events for goodware and ransomware within the dataset is shown in Fig. \ref{fig:events}.

\begin{table}
    \centering
\caption{The goodware and ransomware count used in the dataset to train and test SILRAD}
\label{tab:totalevents}
    \begin{tabular}{|c|c|} \hline 
        \textbf{Event type} & \textbf{Count}\\ \hline 
         Ransomware& 20,710\\ \hline 
 Goodware&176,130\\ \hline
 \textbf{Total}&\textbf{196,840}\\\hline
    \end{tabular} 
\end{table}

\subsection{Metrics evaluation}\label{metrics}
Albeit frequently used, metrics such as the area under receiver operating characteristic (AUC ROC) show high variance when applied to imbalanced datasets \cite{liu2023implications}. The Matthews Correlation Coefficient (MCC), on the other hand, is robust against imbalanced datasets and is considered the preferred metric since a high MCC always produces a high AUC ROC but not necessarily vice versa \cite{chicco2023matthews}. For this reason, we use the MCC to measure the performance of SILRAD and other classifiers throughout the experiment. The MCC is determined using Equation. \ref{eqn:mcc}:

\begin{multline}
    \label{eqn:mcc}
    MCC = \frac{TP \cdot TN - FP \cdot FN}{\sqrt{(TP + FP)\cdot(TP + FN)\cdot(TN + FP)\cdot(TN + FN)}}
\end{multline}

\bigskip

Where \textit{TP} equals true positives, \textit{TN} equates to true negatives, \textit{FP} represents false positives, and finally, \textit{FN} represents the number of false negatives. \\
We also calculate several supporting metrics frequently used throughout the literature, such as accuracy, precision, recall and F1-score, to compare classifiers.    

\bigskip
\begin{align}
    accuracy = \frac{correctly\,classified\,samples}{total\,number\,of\,samples}
\end{align}

\begin{align}
    precision = \frac{TP}{TP + FP}
\end{align}

\begin{align}
    recall = \frac{TP}{TP + FN}
\end{align}

\begin{align}
    F1 score = \frac{2 \times precision \times recall}{precision + recall}
\end{align}

\bigskip

\newpage
\section{Results}\label{sec:results}

We carried out two experiments to validate the effectiveness of SILRAD. First, we measured the performance of traditional ML techniques against the dataset to validate the article's premise. Secondly, we compared the performance of SILRAD with several incremental learning algorithms.

\begin{figure}
    \centering
    \includegraphics[width=.6\linewidth]{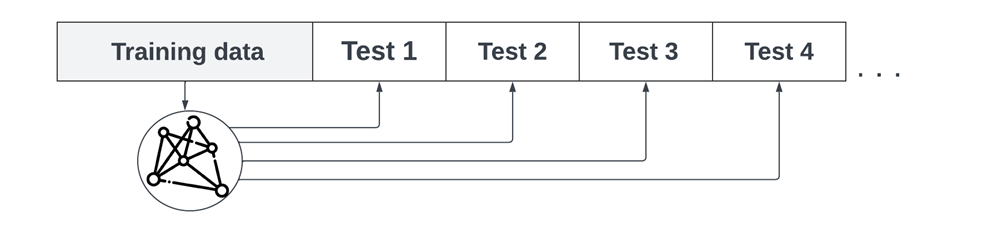}
    \caption{The training and testing approach for traditional ML techniques}
    \label{fig:datablocks}
\end{figure}

\begin{figure*}[t!]
    \centering
    \includegraphics[width=1\textwidth]{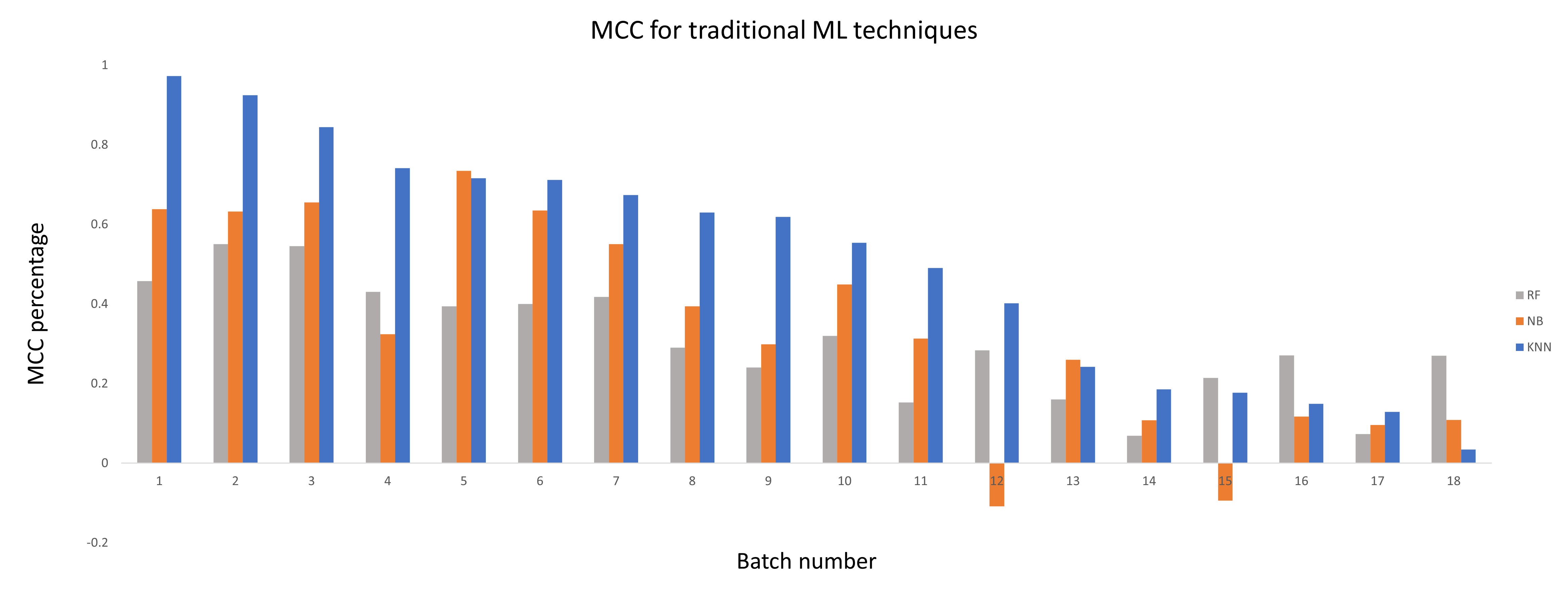}
    \caption{MCC rate of training and testing the dataset using traditional ML techniques}
    \label{fig:traditionalml}
\end{figure*}

\subsection{Experiment one: Measuring the performance of traditional ML techniques with the dataset}\label{sec:traditional}
We first evaluate the performance of popular traditional ML techniques used throughout the literature to detect ransomware activity amongst a continuous stream of Sysmon events. In real-world scenarios, the complete dataset would not be available in totality to classify ransomware behaviour within Sysmon logs. For this reason, we use the Hold-Out method to simulate a continuous stream of data and measure the performance of k-nearest neighbours (KNN), Gaussian Naive Bayes (NB) and Random Forest (RF) on the dataset. This entails splitting the dataset and training and testing on different subsets of data. To simulate concept drift, new ransomware families were gradually introduced throughout the batches. Subsequently, the dataset was split into 20 equal portions, and the ML classifier was trained on the first two blocks of data containing benign and ransomware events.  Then, the model from the first partition was tested on consecutive portions without retraining to observe the performance degradation in the presence of concept drift, as shown in Fig. \ref{fig:datablocks}. For completeness, we test the model 10 times per data segment and calculate the mean to get an accurate representation. \\
It is apparent from Fig. \ref{fig:traditionalml} that the MCC performance suffers heavy degradation when new ransomware is introduced. This outlines the limitations of other proposed ransomware detection techniques utilising traditional ML. Notably, it was observed that some batches reported an MCC rate higher than expected. The authors suggest this is a result of similarities between ransomware families.

\newpage
\subsection{Experiment two: Comparing SILRADs performance with different features}

\begin{table}
    \centering
    \caption{A comparison of the MCC rate for different feature numbers}
    \resizebox{0.60\linewidth}{!}{%
    \begin{tabular}{|c|c|c|c|c|c|} \hline 
         \textbf{Feature} \#&  \textbf{Accuracy}&  \textbf{Precision}&  \textbf{Recall}&  \textbf{F1-Score}& \textbf{MCC}\\ \hline 
         5&  98.88\%&  94.99\%&  94.35\%&  94.67\%& 94.04\%\\ \hline 
         10&  98.83\%&  94.61\%&  94.27\%&  94.44\%& 93.78\%\\ \hline 
         15&  98.81\%&  94.42\%&  94.26\%&  94.34\%& 93.68\%\\ \hline 
         20&  98.84\%&  94.7\%&  94.27\%&  94.49\%& 93.84\%\\ \hline 
         25&  98.79\%&  94.22\%&  94.29\%&  94.26\%& 93.58\%\\ \hline
    \end{tabular}
    \label{tab:feature-numbers}
    }
\end{table}

To optimise SILRAD's performance, we evaluate the classification results using different numbers of features. As highlighted in section \ref{incremental}, the classification engine is underpinned by the ARF model, and to standardise the results, we use a seed value of 42 and Naive Bayes for the leaf prediction. The delta setting for the ADWIN concept drift detection is kept at the default value of 0.002. We test the classification performance with different numbers of features by order of importance as explained in section \ref{selection}. The top 15 features, ranked by importance, are displayed in Figure \ref{fig:features}. Beyond these, additional features contribute insignificantly to detection accuracy. As shown in Table. \ref{tab:feature-numbers}, SILRAD scored the best overall classification performance with an MCC rate of 94.04\% when only 5 of the most significant features were used for classification. This included the TargetObject, task, CallTrace, ParentImage and IntegrityLevel. We observed performance degradation in the MCC when the number of features increased. 

\begin{figure*}[t!]
     \centering
     \begin{subfigure}{0.95\textwidth}
         \centering
         \includegraphics[width=1\textwidth]{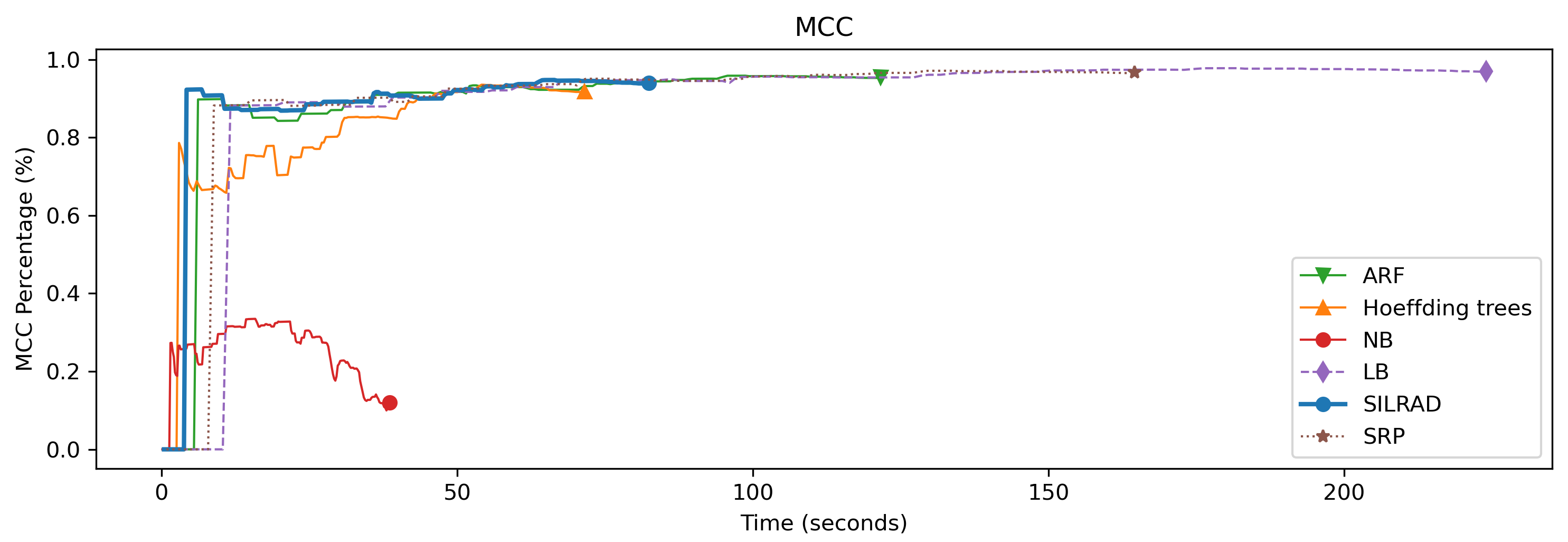}
         \caption{Matthews Correlation Coefficient (MCC) rate over time}
         \label{fig:mcc-graph}
     \end{subfigure}%
     \hfill
     \begin{subfigure}{0.95\textwidth}
         \centering
         \includegraphics[width=1\textwidth]{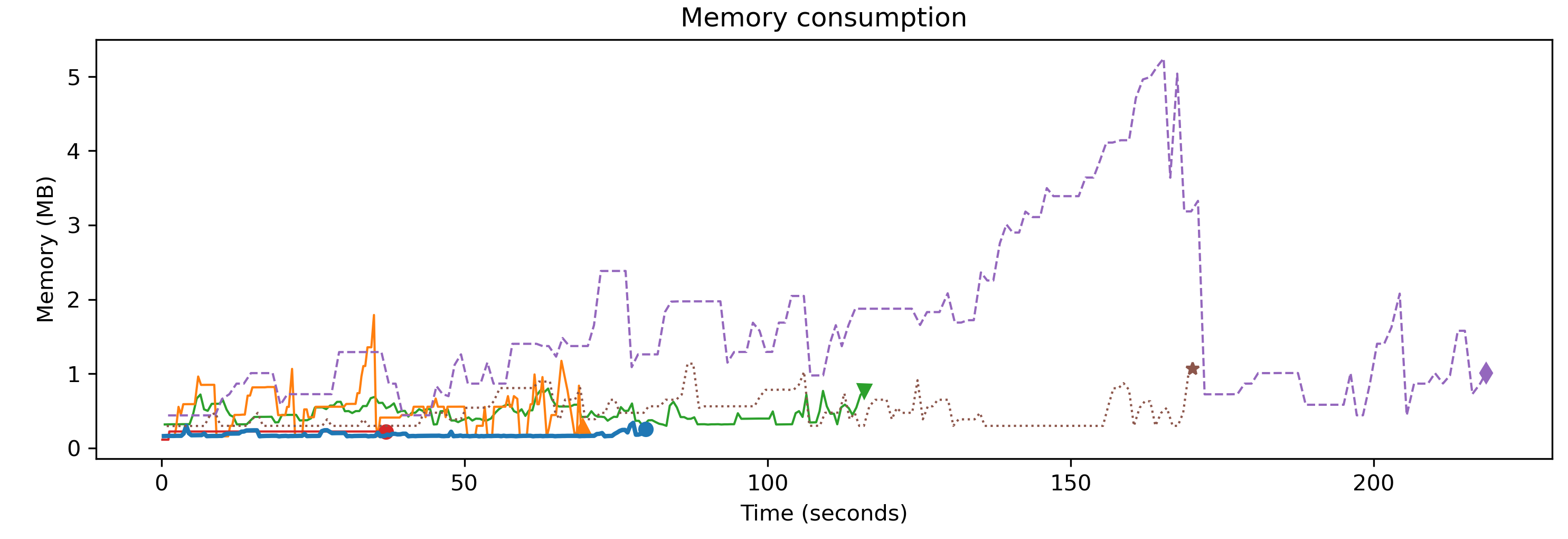}
         \caption{Memory consumption (MB) over time}
         \label{fig:memory-graph}
     \end{subfigure}
        \caption{MCC rate over time and Memory consumption over time for SILRAD and several incremental learning algorithms}
        \label{fig:two-graphs}
\end{figure*}
\newpage
\subsection{Experiment three: Comparing the performance of SILRAD against other incremental learning algorithms}


\begin{table}[]
\centering
\caption{Performance of SILRAD and other ML techniques on Sysmon streams of data}
\label{tab:silrad-results}
\resizebox{0.65\linewidth}{!}{%
\begin{tabular}{|c|c|c|c|c|c|c|}
\hline
\textbf{Classifier} &  \textbf{Incremental}&\textbf{Accuracy} & \textbf{Precision} & \textbf{Recall} & \textbf{F1-Score} & \textbf{MCC} \\ 
\hline 
\rowcolor{yellow!50}
 kNN&  N&88.83\%& 71.61\%& 49.12\%& 51.83\%&51.09\%\\ \hline 
 \rowcolor{yellow!50}
 NB&  N&84.97\%& 61\%& 33.6\%& 36\%&33.95\%\\ \hline 
 \rowcolor{yellow!50}
 RF&  N&90.21\%& 73.32\%& 19.74\%& 26.22\%&30.87\%\\\hline
ARF    &  Y&99.19\% & 96.59\% & 95.67\% & 96.13\% & 95.68\% \\ \hline
HT     &  Y&98.43\% & 90.15\% & 95.46\% & 92.73\% & 91.89\% \\ \hline
NB     &  Y&86.06\% & 24.37\% & 15.43\% & 18.90\% & 12.06\% \\ \hline
LB     &  Y&99.44\% & 97.61\% & 97.03\% & 97.31\% & 97\%    \\ \hline
\textbf{SILRAD} &  \textbf{Y}&\textbf{98.89\%} & \textbf{94.87\%} & \textbf{94.59\%} & \textbf{94.73\%} & \textbf{94.11\%} \\ \hline
SRP    &  Y&99.37\% & 97.03\% & 97.01\% & 97.02\% & 96.67\% \\ \hline
\end{tabular}%
}
\end{table}

\begin{figure}[h!]
    \centering
    \includegraphics[width=0.5\linewidth]{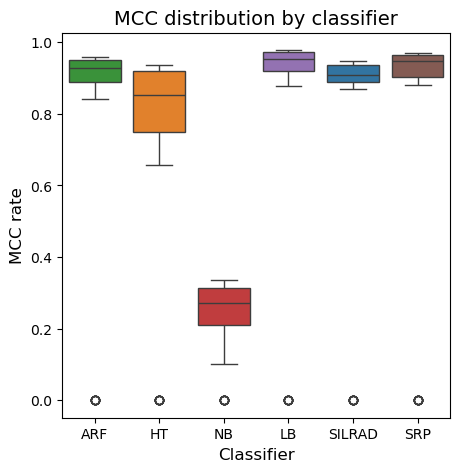}
    \caption{Box plot to show the distribution of the MCC rate for various classifiers}
    \label{fig:boxplots}
\end{figure}

We compare SILRAD's performance against several other traditional ML and online incremental learning techniques, including k-nearest neighbours (kNN), Random Forrest (RF), Adaptive Random Forest (ARF), Hoeffding Trees (HT), Gaussian Naive Bayes (NB) in batch mode and on-line mode, Leveraging Bagging Classifier (LB) and Streaming Random Patches ensemble classifier (SRP). HT was selected as the underpinning model for ensemble models such as LB and SRP, with a grace value of 50 and a delta of 0.00001. To standardise the results, the delta values for the ADWIN drift detection and the grace period were kept at the default values of 0.002 and 10, respectively. 
Traditional ML techniques were trained and tested using the technique explained in section \ref{sec:traditional}. It is clear from the results presented in Table. \ref{tab:silrad-results} that the overall detection accuracy of traditional ML techniques is significantly lower than incremental learning techniques. For incremental learning techniques, benign and ransomware events were ingested into each classifier as a continuous stream and evaluated. \\We observe that SILRAD yielded 98.88\% accuracy, 94.99\% precision, 94.35\% recall, an F1-score of 94.67\% and a Matthews correlation coefficient (MCC) 94.04\%. Importantly, it was observed that SILRAD's performance improved over time as new ransomware was introduced into the data stream, as shown in Fig. \ref{fig:mcc-graph}, solidifying ADWIN's usefulness for concept drift detection. Although the overall MCC performance is slightly less than other incremental classifiers, SILRAD consumes fewer resources and is faster than other models, making it advantageous when detecting ransomware in real-time. \\
Although more accurate, classification with SRP and LB took considerably longer than SILRAD. Conversely, HT completed classification faster but displayed significant variance in the MCC rate and required nearly the whole dataset to achieve an MCC rate of over 90\%. Like the classification time, SRP and LB consumed the most resources, as shown in  Fig. \ref{fig:memory-graph}. SILRAD consumed significantly less memory than other classifiers, achieving an MCC rate of over 90\% and displayed the least variance in the MCC value, as shown in Fig. \ref{fig:boxplots}.
 \\
Overall, SILRAD proved its ability to detect ransomware accurately faster than other incremental learning algorithms whilst maintaining a minimal resource footprint. 

\newpage
\section{Conclusion and future work}
This paper proposed the Sysmon Incremental Learning system for Ransomware Analysis and Detection (SILRAD) technique to detect ransomware activity. Using only five features derived from Sysmon logs, we could detect ransomware attacks from AvosLocker, BlackBastsa, Conti, Hive, Lockbit, and REvil with a 98.89\% accuracy rate, 94.87\% precision, 94.59\% recall, 94.73\% F1-score and an MCC rate of 94.11\%. SILRAD's concept drift detection automatically detected drift within the dataset. Despite operating within a heavily imbalanced dataset, SILRAD improved its detection accuracy over time, allowing it to detect new ransomware samples as they arrive without retraining the model from scratch. By contrast, other non-incremental learning approaches popularised throughout the literature degraded over time as new ransomware was introduced. \\
Although the final MCC rate of SILRAD was not as high as other incremental learning algorithms, its low memory consumption, low variance, and quick detection time make SILRAD a suitable candidate for real-time ransomware detection.
Our approach highlights SILRAD's utility at quickly and accurately detecting new ransomware within an evolving datastream whilst ensuring that sensitive data is not vulnerable between model updates. Future work will focus on improving the detection accuracy of SILRAD.

\section{Acknowledgement}
The work has been supported by the Cyber Security Research
Centre Limited, whose activities are partially funded
by the Australian Government’s Cooperative Research Centres
Programme.

\bibliographystyle{elsarticle-num}
\bibliography{references} 

\newpage
\section{Biography Section}
\bio{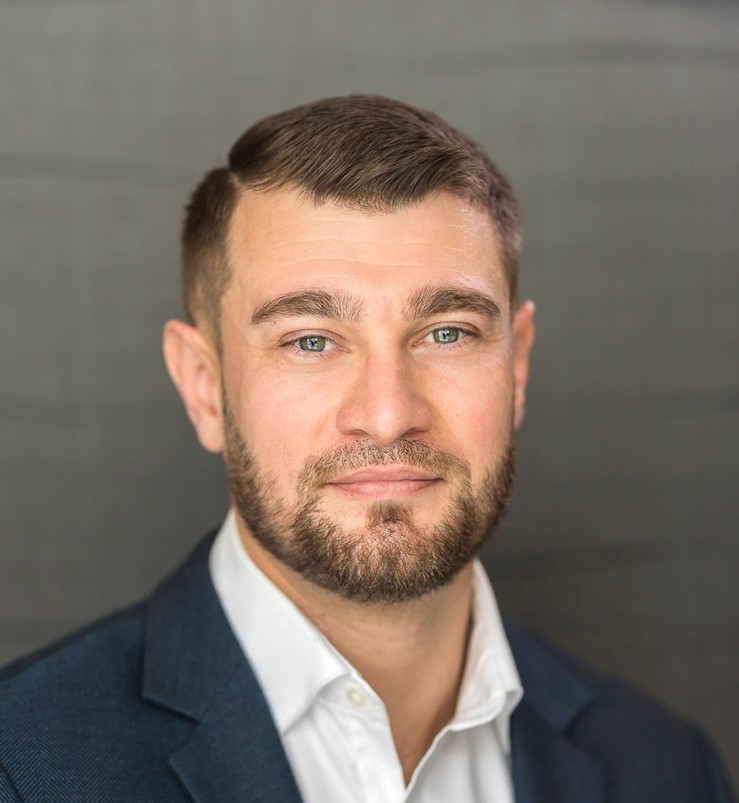}\textbf{Jamil Ispahany} is pursuing a PhD at the School of Computing, Mathematics and Engineering at Charles Sturt University, Australia. He is a recipient of a scholarship at the Cyber Security Cooperative Research Centre (CSCRC). His research interests include cyber security, machine learning and malware detection.
\\
\\
\\
\\
\\
\endbio
\bio{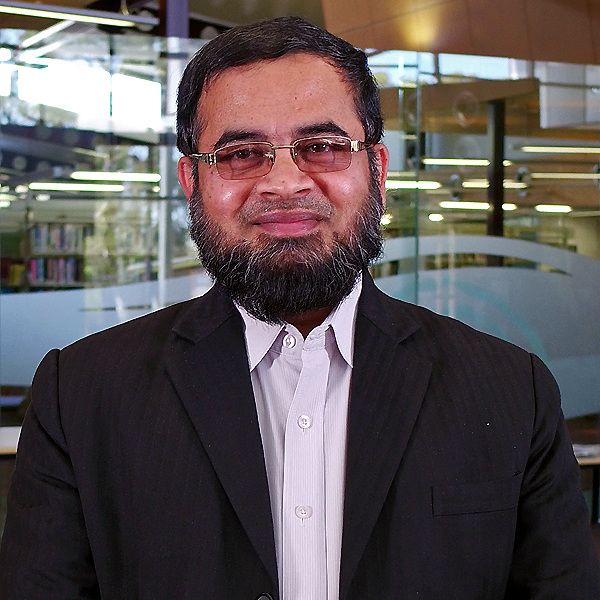}
\textbf{Md Rafiqul Islam} is working as an Associate Professor at the School of Computing, Mathematics and Engineering, Faculty of Business, Justice and Behavioral Sciences, Charles Sturt University, Australia. Dr Islam has a strong research background in Cybersecurity with a specific focus on malware analysis and classification, Authentication, security in the cloud, privacy in social media and the Internet of Things (IoT). He is leading the Cybersecurity research team and has developed a strong background in leadership, sustainability, and collaborative research. He has a strong publication record and has published more than 180 peer-reviewed research papers, book chapters and books. His contribution is recognised both nationally and internationally by achieving various rewards such as professional excellence, research excellence, and leadership awards.
\endbio

\bio{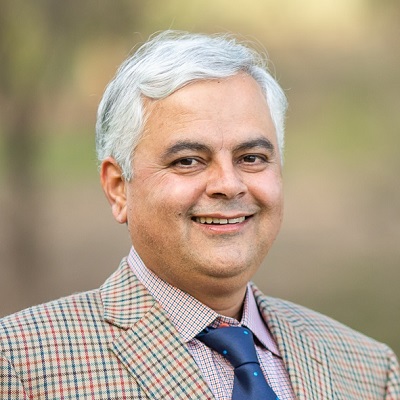}
\textbf{M. Arif Khan} received the B.Sc. degree in electrical engineering from the University of Engineering and Technology Lahore, Pakistan, the M.S. degree in electronic engineering from the GIK Institute of Engineering Sciences and Technology, Pakistan, and the Ph.D. degree in electronic engineering from Macquarie University Sydney, Australia. He is currently a Senior Lecturer with the School of Computing, Mathematics and Engineering, Charles Sturt University, Australia. His research interests include future wireless communication technologies, smart cities, massive MIMO systems, and cyber security. He was a recipient of the Prestigious International Macquarie University Research Scholarship (iMURS), and ICT CSIRO scholarships for his Ph.D. degree. He also has the competitive GIK Scholarship for his master’s degree.
\endbio

\bio{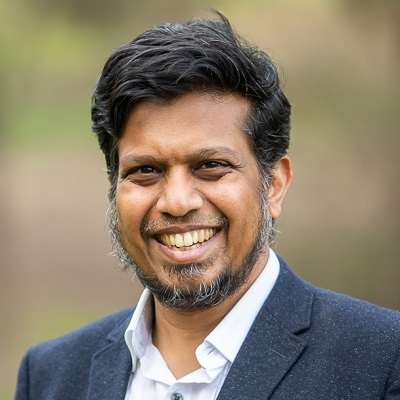}
\textbf{Md Zahidul Islam} (known as Zahid Islam) is a Professor of Computer Science, in the School of Computing, Mathematics and Engineering, Charles Sturt University, Australia. His main research interests are in Data Mining, Knowledge Discovery, Privacy Preserving Data Mining, and Applications of Data Mining/Machine Learning in various areas, including Cyber Security. URL:http://csusap.csu.edu.au/zislam/
\endbio

\end{document}